\documentclass{article}

\usepackage{arxiv}

\usepackage[utf8]{inputenc} % allow utf-8 input
\usepackage[T1]{fontenc}    % use 8-bit T1 fonts
\usepackage{hyperref}       % hyperlinks
\usepackage{url}            % simple URL typesetting
\usepackage{booktabs}       % professional-quality tables
\usepackage{amsfonts}       % blackboard math symbols
\usepackage{nicefrac}       % compact symbols for 1/2, etc.
\usepackage{microtype}      % microtypography
\usepackage{lipsum}		    % Can be removed after putting your text content
\usepackage{graphicx}
\usepackage{subcaption}
\graphicspath{ {./images/} }
\usepackage{natbib}
\usepackage{doi}
\usepackage{enumitem}

\newtheorem{definition}{Definition}

\title{Reputation analysis of news sources in Twitter: Particular case of Spanish presidential election in 2019}

\date{} 					% Or removing it

\author{ \href{https://orcid.org/0000-0001-8332-0381}{\includegraphics[scale=0.1]{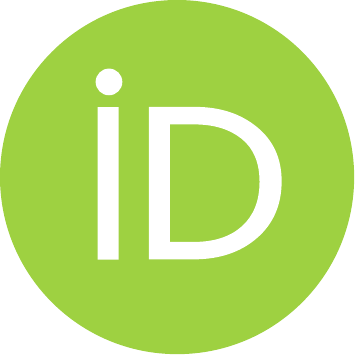}\hspace{1mm}Aarón~López-García}\\
	Department of Computer Science\\
	University of Valencia\\
	\texttt{logara8@alumni.es}\\
	\And
	\href{https://orcid.org/0000-0000-0000-0000}{\includegraphics[scale=0.1]{orcid.pdf}\hspace{1mm}Rafael~Benítez}\\
	Department of Business Mathematics\\
	University of Valencia\\
	\texttt{rafael.suarez@uv.es}
}

\renewcommand{\undertitle}{}
\renewcommand{\shorttitle}{Reputation analysis of news sources in Twitter}

\hypersetup{
pdftitle={Reputation analysis of news sources in Twitter: Particular case of Spanish presidential election in 2019},
pdfsubject={stat.ME, stat.AP, math.DS},
pdfauthor={Aarón~López-García, Rafael~Benítez},
pdfkeywords={Reputation, Twitter, Spain},
}

\begin{document}
\maketitle

\begin{abstract}
Fake news are affecting a large proportion of the population even becoming a danger to the society. Mostly, this disinformation flow take place through Internet. Being aware of that problem, in this work we propose a synthetic indicator that measures the user reputation in Twitter in order to analyze the credibility of the content in this social network. In order to show the indicator utility, we have analyzed data from some political topics in Spain from 2019 to 2020 and we have checked that bots plays a decisive role into the spread of news and, as might be expected, the link among popularity and reputation reports about the event credibility.
\end{abstract}

% keywords can be removed
\keywords{Reputation \and Twitter \and Spain \and Graph \and Indicators}

\section{Introduction}\label{sec:intro}

Reputation is understood as the opinion that people in general have about someone or something, or how much respect or admiration someone or something receives, based on past behaviour or character. It is obvious that such term is subjective and it can vary depending on the concept, society or culture. In this paper, we want to focus on Online Social Networks (OSN) in which fake news can play a key role due to the duality of misinformation and trustworthiness. Possibly, the most evident case of propaganda in social networks and monetary interests, is the case of Cambridge Analytica founded in 2003 and forced to break up in 2018 due to the scandal of Brexit and the USA campaign both in 2016. Nowadays, with the raise of the technologies 4.0, false information is taking over every digital platform. Moreover, modern artificial intelligence techniques can be applied in relevant networks as Facebook and Twitter in order to mislead users and create hoaxes and spread them so fast. In regard of the accounts, we would like to highlight the role of bots, that is to say, accounts digitally created and monitored that create, disseminate and lend credibility to fake news.

\section{Related work}\label{sec:related_work}

The study of social media content has been faced from different approaches, in which we can distinguish: User reputation \citep{Castillo2011, Alrubaian2017}, Content credibility \citep{Yang2019}, Timeline rank \citep{Bhowmick2019}, and Fake news detection \citep{Gao2018}.

In this paper we have followed the user reputation framework based on our last work \citep{Aaron2019}. Now, our goal is to generate a set of indicators that, all together, can measure the reputation of a Twitter account. To this effect, we have tracked and analyzed Twitter content related with the Spanish presidential election conducted in November 2019, from 1/9/2019 to 1/2/2020.

\section{Methodology}\label{sec:methodology}

We have developed various tools from the different sources mentioned in section \ref{sec:related_work}, hence we split this section in order to clarify the methods implemented. The novelty designed is focused on the subsection \ref{subsec:reputation} and the other subsections have been detailed to understand not only the definition of our indicator but the perspective as well.

\subsection{Graph analysis}\label{subsec:graph_analysis}

Twitter's structure allows us to relation the users with a graph $G=(V, E)$, in which $V$ denotes the vertices (accounts) and $E$ the relationships (edges). We manage such links depending upon the problem, in any case the vertices are the accounts. For users approach, we represent the graph as $G_U$ in which $E_U$ describes the friends-follower relation. For tweets approach, we denote $G_T$ as the tree graph with the timeline of the event, so each vertex represent the tweet-retweet messages. In both cases, we can mathematically formulate definitions, metrics and functions to describe the behaviour in Twitter.

\begin{definition}\label{def_notions}
Let $u$ and $v$ be two accounts in a set of vertices $V$, we name:
\begin{itemize}
    \item Path: Each $n$-tuple defined in $G$ that connects two vertices, i.e., $P_{uv}=(e_1,...,e_n)$ where $\{e_1,...,e_n\}\subset E$ with $e_1=u$ and $e_n=v$ and $\overline{e_i e_{i+1}}\in V$.
    \item Adjacency matrix: Given a graph $G=(V, E)$, the adjacency matrix associated is the binary matrix $M_{Adj}=A_{uv}$ in which their values are $1$ if $\overline{uv}\in V$ and $0$ otherwise.
    \item Cardinality: It is the total amount of elements in a set. Given a set $X$, such application is denoted as $|X|$.
\end{itemize}
\end{definition}

\begin{definition}[Node scores]\label{def_node_scores}
Let $u$ and $v$ be two accounts in a set of vertices $V$, we assume that our graph is finite, so $V$ has a finite number of elements $|V|=n<\infty$:
\begin{enumerate}[label=\roman*.]
    \item\label{def_distance} Distance: Minimal path that joins two nodes, $\displaystyle d(u,v)=\min_{u,v\in V}\left\{|P_{uv}|:P_{uv}\subset E_U\right\}$.
    \item\label{def_diagonal} Diagonal: Maximal from one node to another, $\displaystyle Diag(G_U)=\max_{u,v\in V}d(u,v)$.
    \item\label{def_degree} Node degree: $\chi(v)=\chi_{I}(v)+\chi_{O}(v)$, being $\chi_{I}$ and $\chi_{O}$ the input and output degree respectively whose definition is:
    $$\chi_{out}(v) = \displaystyle\sum_{w\in V}A_{vw}\qquad and \qquad\chi_{in}(v) =\sum_{w\in V}A_{wv}.$$
    \item\label{def_closeness} Closeness: $\displaystyle C(v)=\sum_{w\neq v}\frac{1}{d(v,w)}$.
    \item\label{def_betweenness} Betweenness: $\displaystyle B(v)=\sum_{u\neq v\neq w}\frac{\sigma_{uw}(v)}{\sigma_{uw}}$, where $\sigma_{uw}$ is the number of shorter paths in $G$ from $u$ to $w$ according to the distance $d$ and $\sigma_{uw}(v)$ is the same than $\sigma_{uw}$ but containing the $v$ vertex.
    \item\label{def_PR} PageRank: $\displaystyle PR(v)=(1-\alpha)\sum_{k\ge 0}\alpha^k M^k e_n$, being $\alpha\in ]0,1[$, $e_n$ the unit vector at $n^{th}$ positon, and $M$ defined as
    \begin{equation}
        M=(M_{vw})_{v,w\in V_U}=\left\{ \begin{array}{lcc}
            A_{vw}/\chi_{out}(w) &   si  & \chi_{out}(w)\ge 0 \\
            1/n &  si & \chi_{out}(w)= 0
                 \end{array}
            \right.
    \end{equation}
\end{enumerate}
\end{definition}

We can utilize formulas \ref{def_distance} and \ref{def_degree} to quantify the influence in a network. For centrality and connectivity, formulas \ref{def_closeness}, \ref{def_betweenness} and \ref{def_PR} can picture it with different insights. Finally, formula \ref{def_diagonal} can be used to normalize distances or to contrast with the notions \ref{def_notions}.

\subsection{Bot scores}\label{subsec:bot}

In order to spot the fake accounts, it is essential to determine an indicator based on the probability to be a bot. Hence, we need to deploy a set of indicators that complement such idea. With the same approach as \citep{davis2016}, we have considered five different indicators which are: User ($a_u$), Sentiment ($a_s$), Content ($a_c$), Temporality ($a_t$), and Bot ($a_b$). Being more specific with the bot feature, Table \ref{table:baremo_bot} has the tag-scale of the accounts in regard of such variable.

\begin{table}
\caption{Scale of the bot score}
\centering
\begin{tabular}{p{0.25\textwidth}p{0.25\textwidth}}
  \toprule
   Tag      & Scale \\ 
  \midrule
   Bot      & $0.0\le a_b\le 0.2$ \\ 
   Doubtful & $0.2  < a_b\le 0.5$ \\ 
   Medium   & $0.5  < a_b\le 0.8$ \\ 
   Real     & $0.8  < a_b\le 1.0$ \\ 
   \bottomrule
\end{tabular}
\label{table:baremo_bot}
\end{table}

\subsection{Reputation scores}\label{subsec:reputation}

The aim of classifying the accounts as reputed is the filtered of Twitter content. The underlying idea of the proposed indicator rely on human behaviour in the network. Our proposal is based on the use of a synthetic indicator that measures and describes the reputation of each individual. Such indicator depends on both active and passive participation in the network. Let $v$ be a Twitter account with network $N^v=N_{I}^v\cup N_{O}^v$ divided as friends and followers users (Input/Output relationship) and $\{a_u, a_s, a_c, a_t, a_b\}$ its account features, we define:

\begin{itemize}
    \item Active reputation
    \begin{equation}\label{eq:rep.activa}
        R_A(v)=
        \big|\big|(a_u, a_s, a_c, a_t, a_b)\big|\big|_G=
        \sqrt[5]{a_u\cdot a_s\cdot a_c\cdot a_t\cdot a_b}.
    \end{equation}

    \item Passive reputation
    \begin{equation}\label{eq:rep.pasiva}
        R_P(v)=
        \frac{\omega_I}{|N_{I}^v|}\sum_{u\in N_{I}^v}R_A(u)+
        \frac{\omega_O}{|N_{O}^v|}\sum_{u\in N_{O}^v}R_A(u)=
        \omega_I R_P^{I}(v) + \omega_O R_P^{O}(v).
    \end{equation}

    \item Total reputation
    \begin{equation}\label{eq:rep.total}
        R(v)=\gamma R_A(v)+ (1-\gamma)R_P(v).
    \end{equation}
\end{itemize}

The parameters $\omega_I$ and $\omega_O$ define a linear convex combination between the reputation of the different nets. The $\gamma$ parameter works similarly between $R_A$ and $R_P$. These parameters allow us to modelize Eq.\ref{eq:rep.total} by following the graph structure defined in \ref{subsec:graph_analysis} and describing the notion in \ref{subsec:bot}. Since we will emphasize with the relation of user content and its reputation, Table \ref{table:baremo_Rep_Activa} shows the scales that we have considered.

\begin{table}[h!]
\caption{Active reputation scale}
\centering
\begin{tabular}{p{0.25\textwidth}p{0.25\textwidth}}
  \toprule
   Tag           & Scale \\ 
  \midrule
   Non-reputed   & $0.00\le R_A\le 0.35$ \\ 
   Doubtful      & $0.35 <  R_A\le 0.50$ \\ 
   Trustworthy   & $0.50 <  R_A\le 0.85$ \\ 
   Reputed       & $0.85 <  R_A\le 1.00$ \\ 
  \bottomrule
\end{tabular}
\label{table:baremo_Rep_Activa}
\end{table}

\subsection{Data}\label{subsec:data}

The data has been obtained through the Twitter API since 1/9/2019 to 1/2/2020 because the political campaigns started on September. On the one hand, we have collected data via R with the use of the Rtweet package \citep{Rtweet2019}. On the other hand, the account features have been computed with the Botometer Python package \citep{botometer2019}, in which it is necessary to have a RapidAPI key as interface. For a better understanding of the methods that has been mentioned, the reader can check our GitHub repository \url{https://github.com/Aaron-AALG/Analisis-de-la-reputacion-en-Twitter}.

\section{Results}\label{sec:results}

We want to study the Spanish presidential election that was conducted on October 10 of 2019. With this aim, we have selected five random accounts that can be labeled as bots according to our criteria (Table \ref{table:baremo_bot}) and the five main parties in Spain in those days. For an in depth research, we have analyzed the account features, relationships graph and messages graph.

\subsection{Features analysis}\label{subsec:features_analysis}

\begin{figure}[t!]
\centering
    \begin{subfigure}{.45\textwidth}
      \centering
      \includegraphics[width=\linewidth]{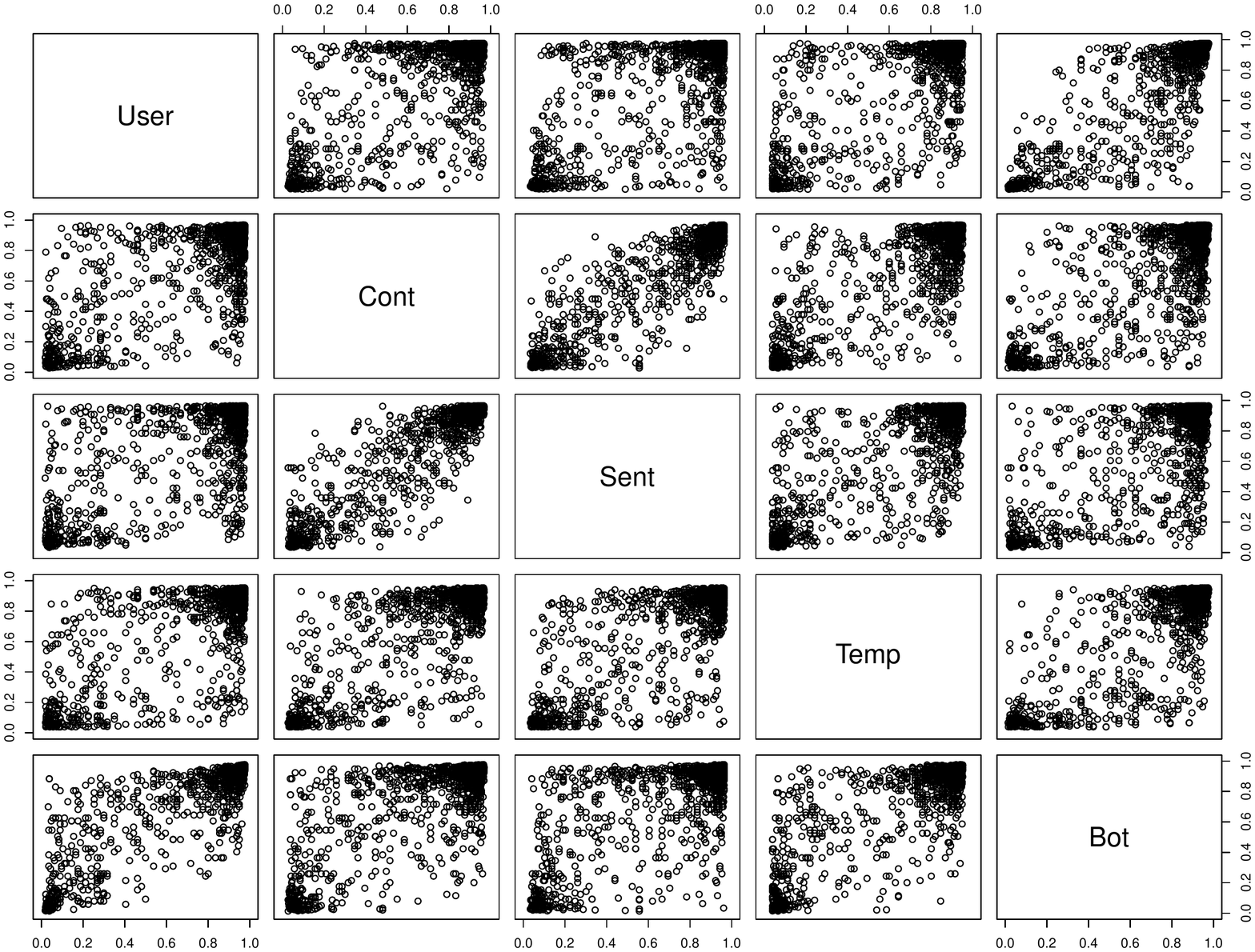}
      \caption{Scatter-plot of the account features}
      \label{fig:features_a}
    \end{subfigure}
    \begin{subfigure}{.45\textwidth}
      \centering
      \includegraphics[width=.9\linewidth]{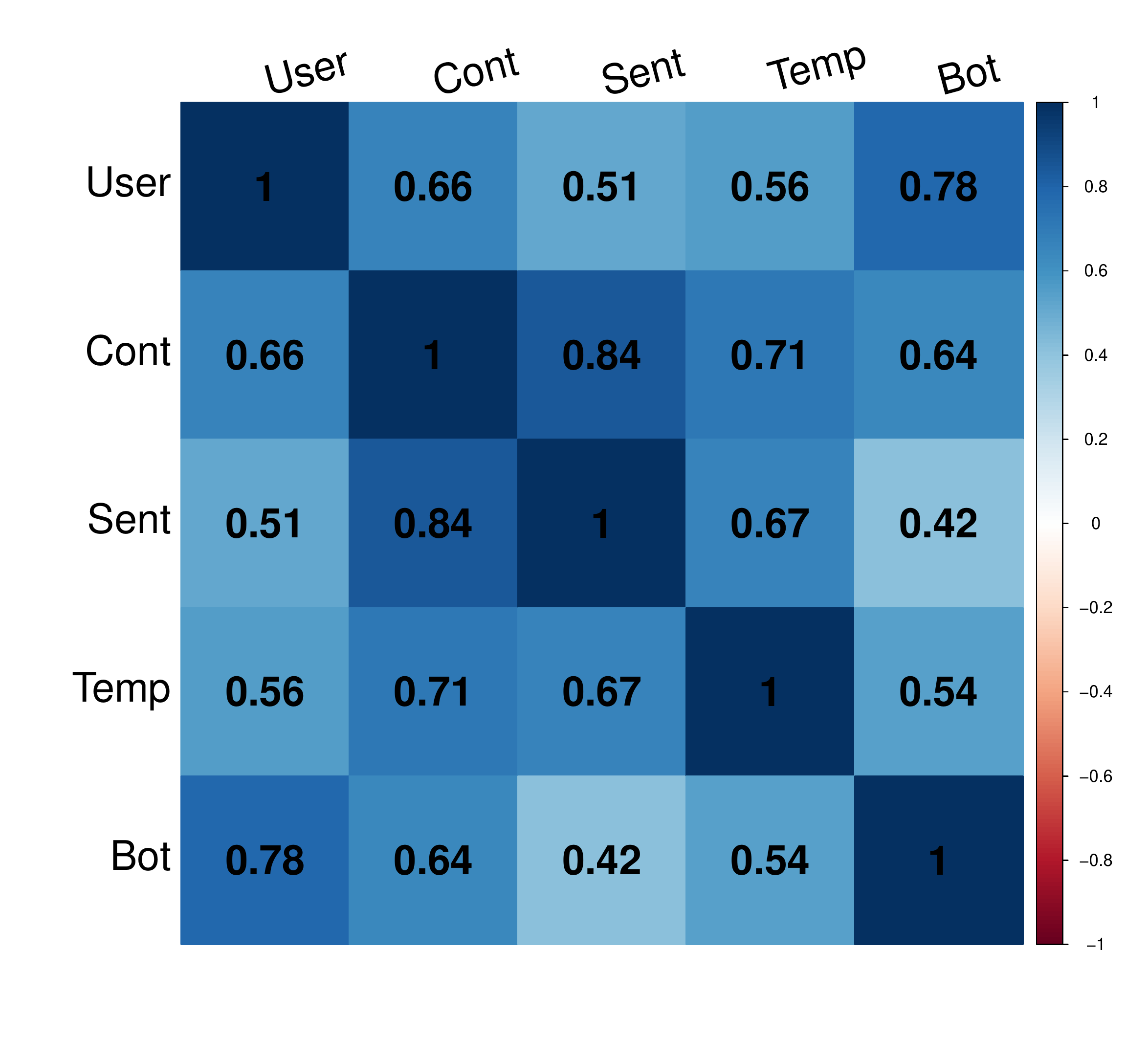}
      \caption{Correlation matrix}
      \label{fig:features_b}
    \end{subfigure}
\caption{Properties of the account features over the tenth part of our dataset.}
\label{fig:features}
\end{figure}

When studying the resemblance among the selected features, we can detect in subfigure \ref{fig:features_a} a pair wise direct relation in the neighborhoods of $(0,0)$ and $(1,1)$. That is why subfigure \ref{fig:features_b} only has positive values. In any case, the figure \ref{fig:features} depicts the inner similarities.

\subsection{Accounts graph}\label{subsec:relationship_graph}

\begin{figure}[b!]
\centering
    \begin{subfigure}{.45\textwidth}
      \centering
      \includegraphics[width=\linewidth]{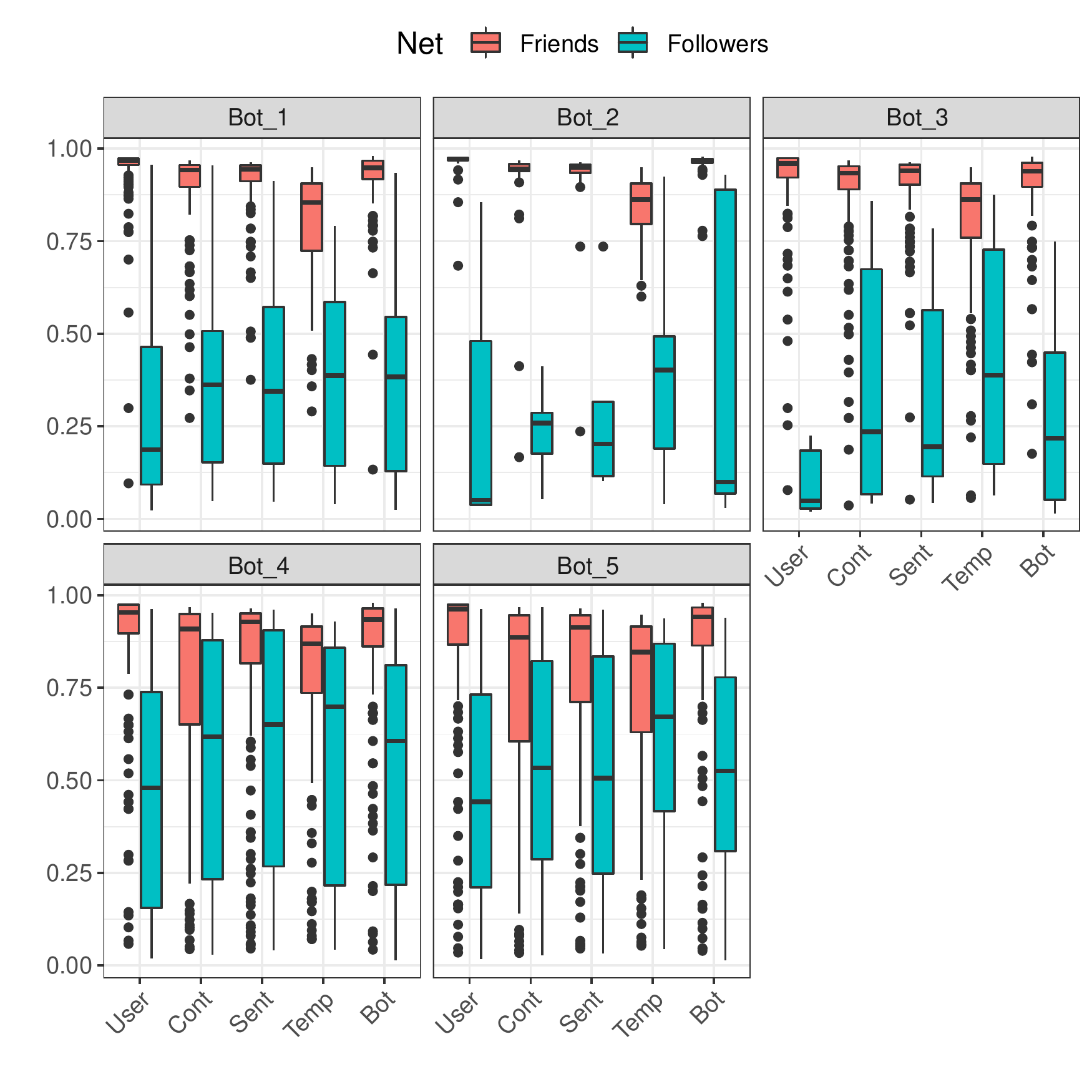}
      \caption{Bots sample}
      \label{fig:example_a}
    \end{subfigure}
    \begin{subfigure}{.45\textwidth}
      \centering
      \includegraphics[width=\linewidth]{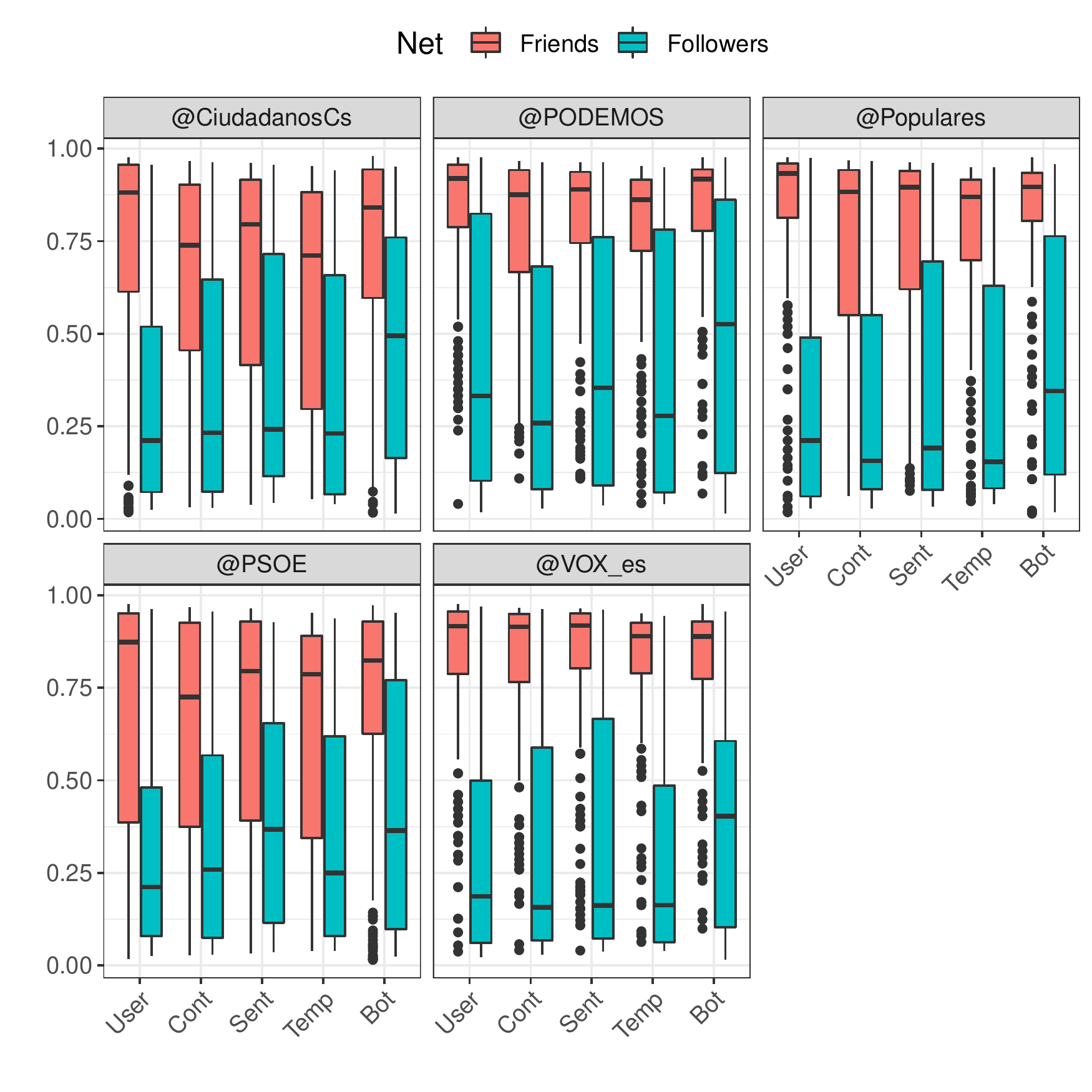}
      \caption{Main parties sample}
      \label{fig:example_b}
    \end{subfigure}
\caption{Distribution of the account features regarding the different nets}
\label{fig:example}
\end{figure}

Since the flux of information goes through the net with clear input-output direction, we can study the difference between the friends and following users per each account. Figure \ref{fig:example} depicts both nets, showing an evident gap between $N_I$ and $N_O$. Moreover, the sample selected has accounts very distinguished and so it is demonstrated in subfigures \ref{fig:example_a} and \ref{fig:example_b}.

The proposed reputation indicators have been computed by fixing $\omega_I=0.75$, $\omega_O=0.25$, and $\gamma=0.7$ as expert decision adequate to the Twitter data. Thus, it is easy to understand that we have given special attention to the account activity within the network because we consider that passive activity is harder to manage in terms of user maintenance. Regarding account connections, we have decided to add a higher weight to the friends net since the graph designed of Twitter sends the information from friends towards followers.

\begin{figure}[t!]
\centering
    \begin{subfigure}{.19\textwidth}
      \centering
      \includegraphics[width=\linewidth]{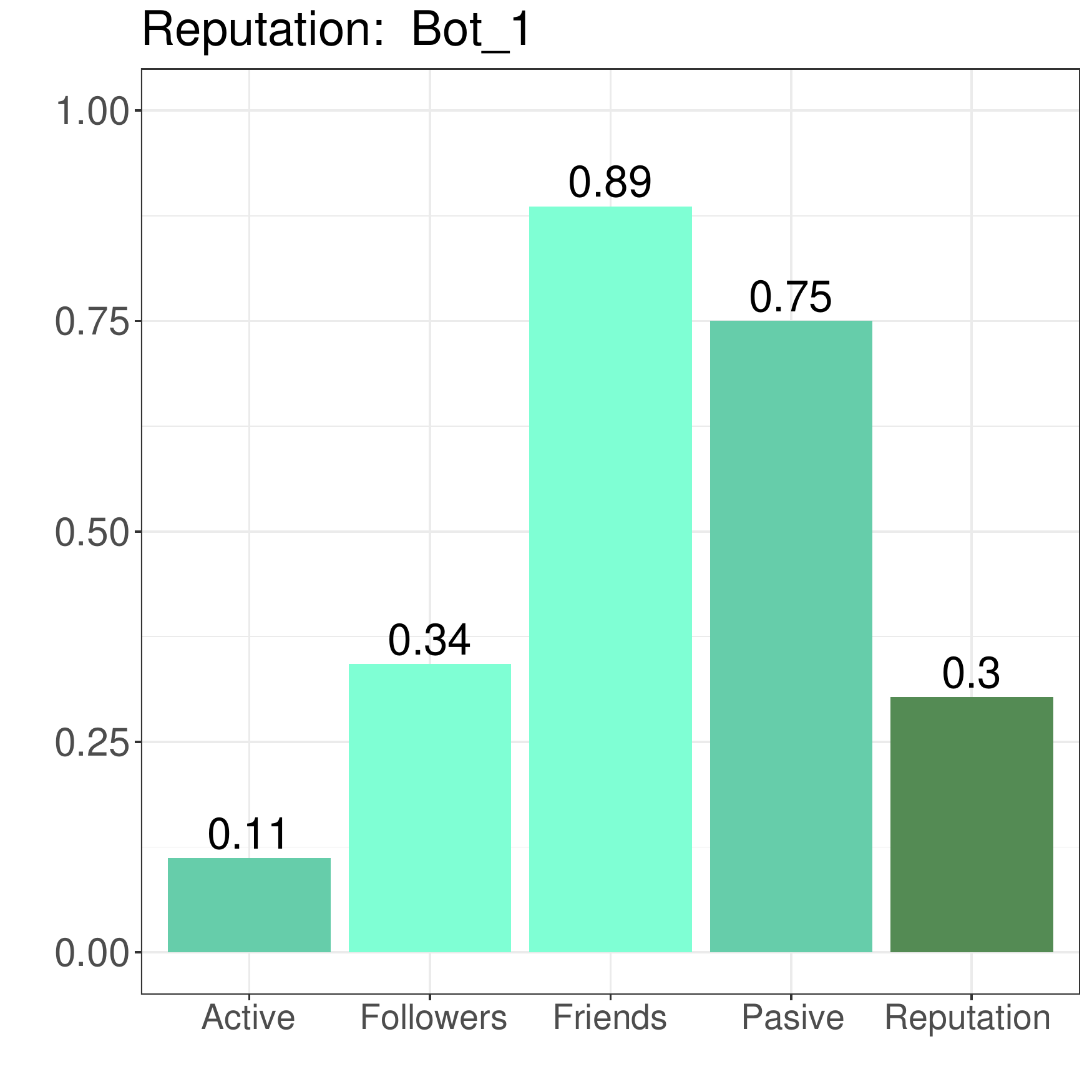}
    \end{subfigure}
    \begin{subfigure}{.19\textwidth}
      \centering
      \includegraphics[width=\linewidth]{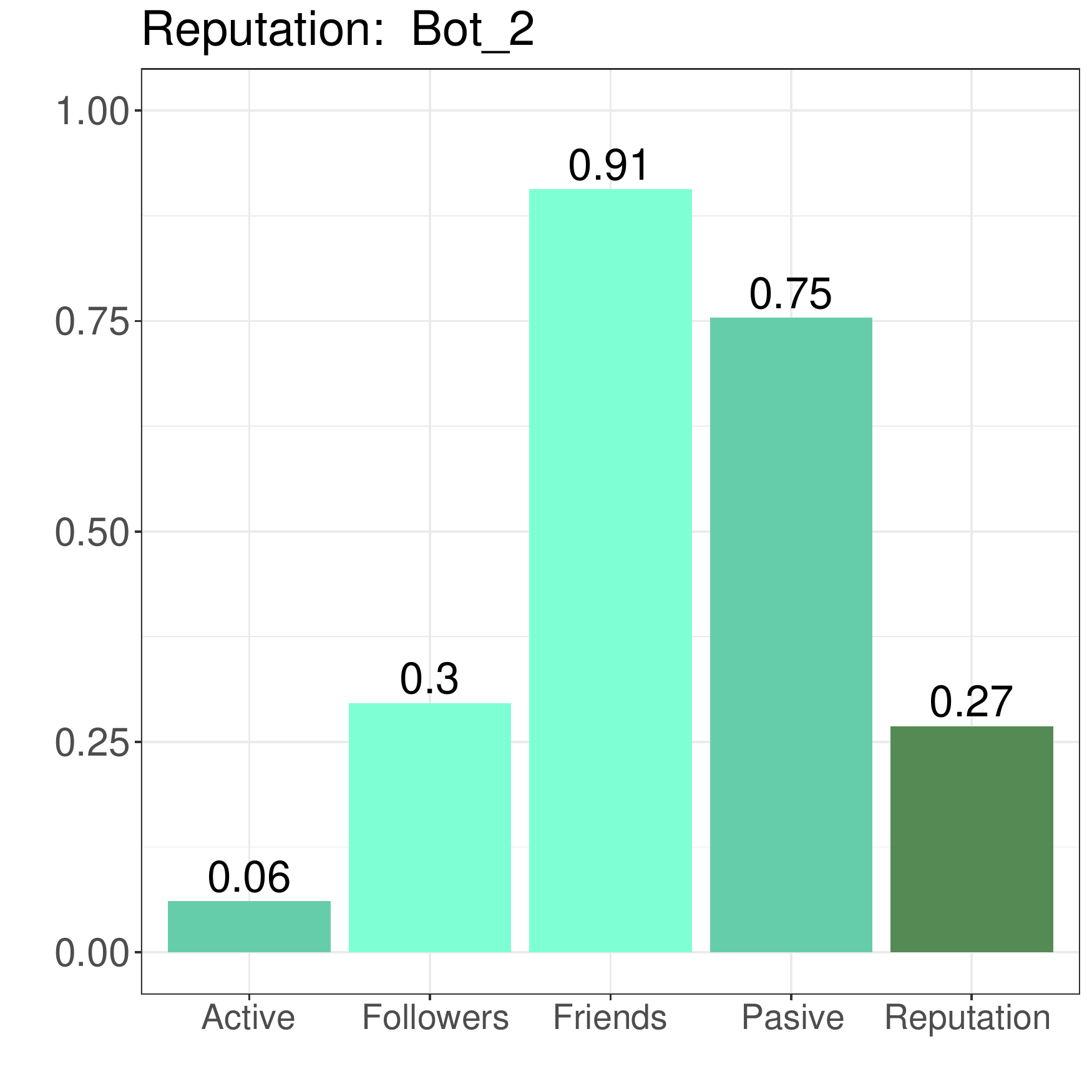} 
    \end{subfigure}
    \begin{subfigure}{.19\textwidth}
      \centering
      \includegraphics[width=\linewidth]{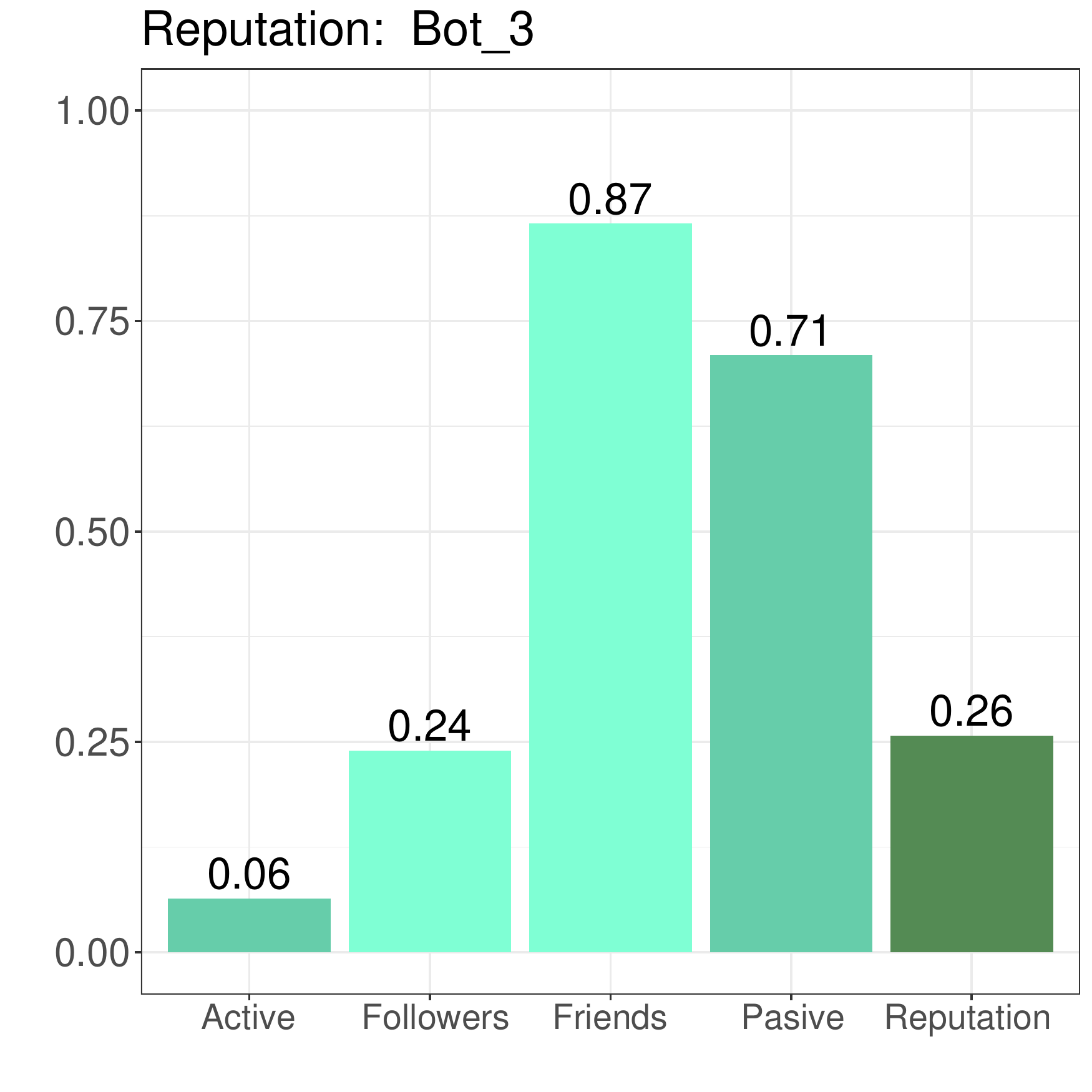}  
    \end{subfigure}
    \begin{subfigure}{.19\textwidth}
      \centering
      \includegraphics[width=\linewidth]{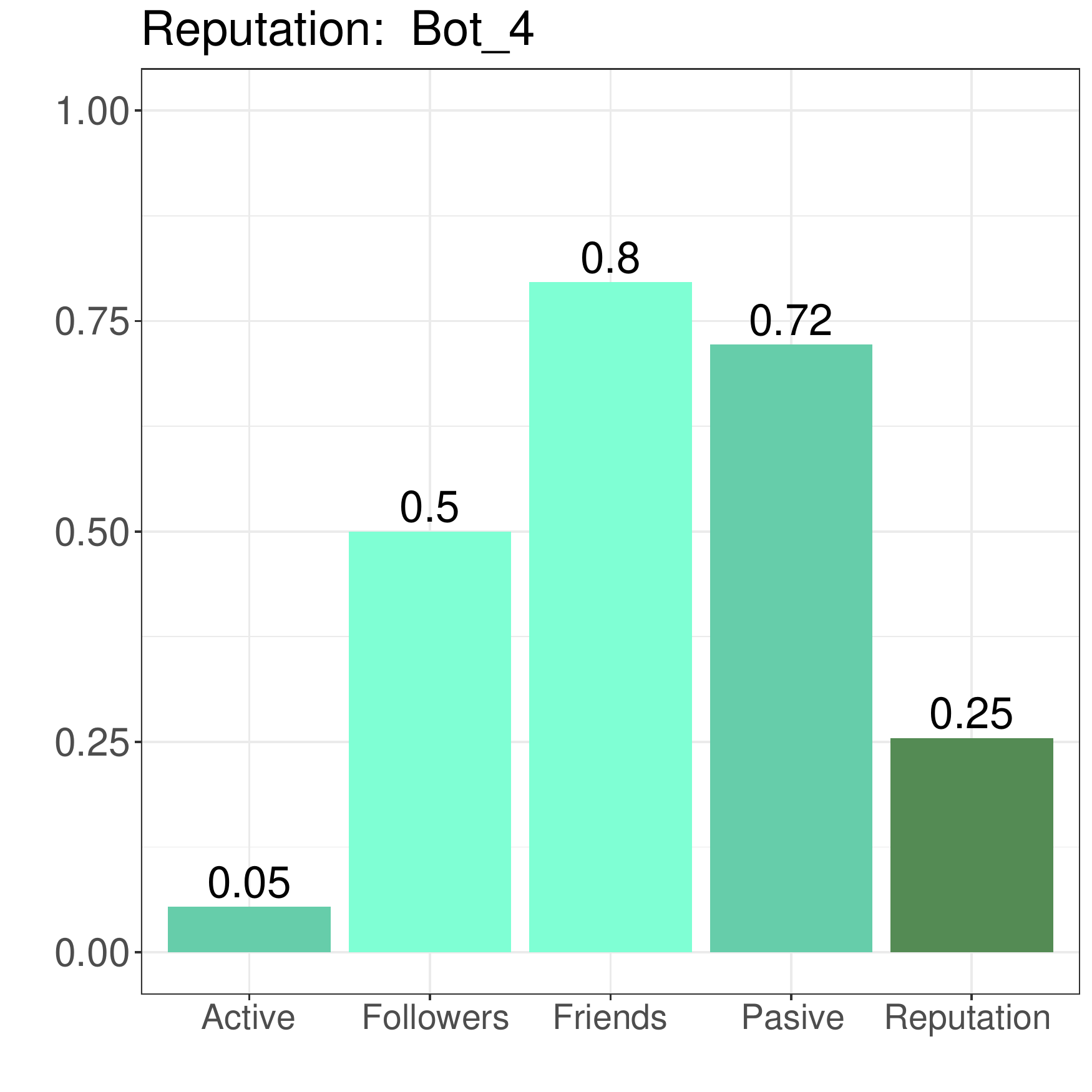}
    \end{subfigure}
    \begin{subfigure}{.19\textwidth}
      \centering
      \includegraphics[width=\linewidth]{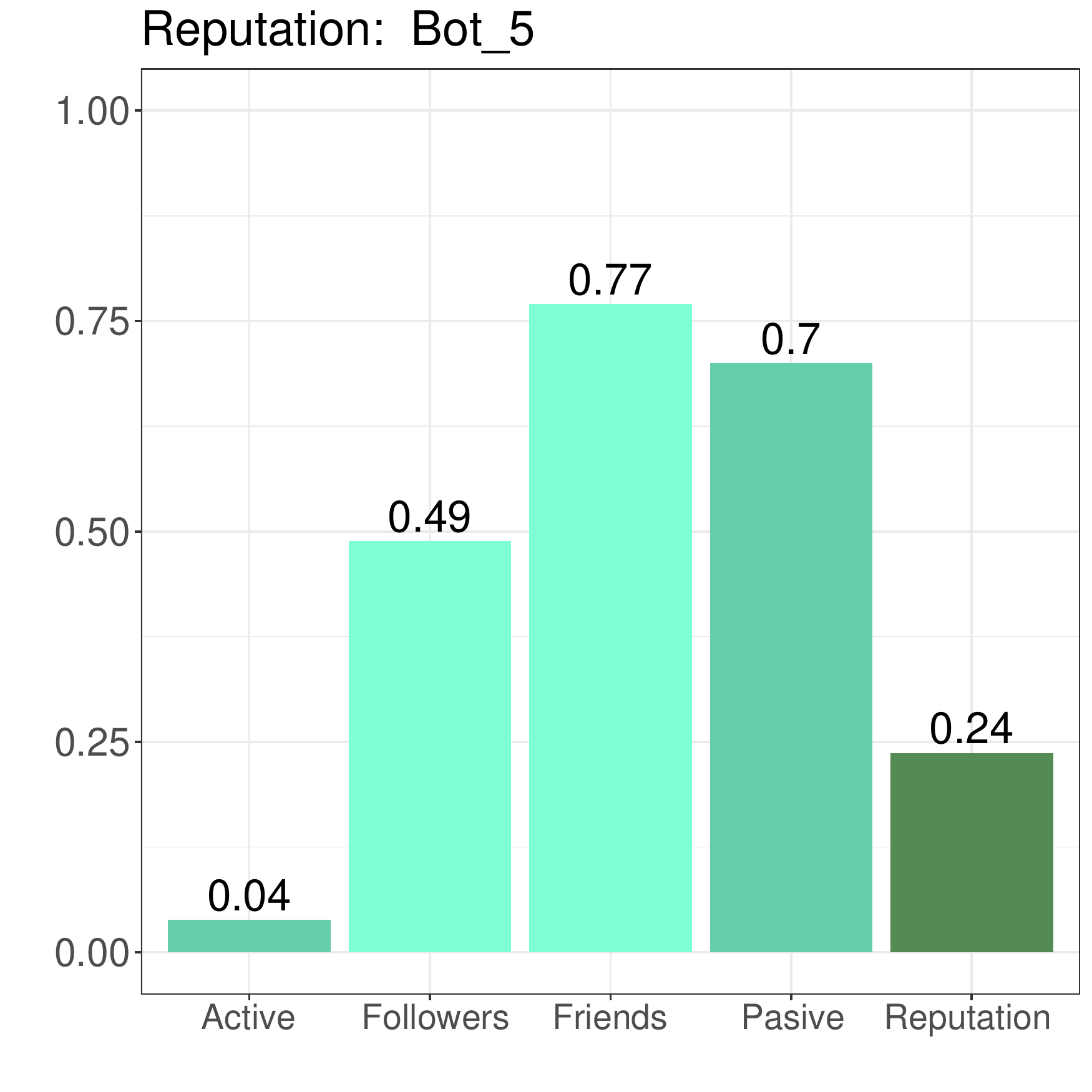} 
    \end{subfigure}
    
    \begin{subfigure}{.19\textwidth}
      \centering
      \includegraphics[width=\linewidth]{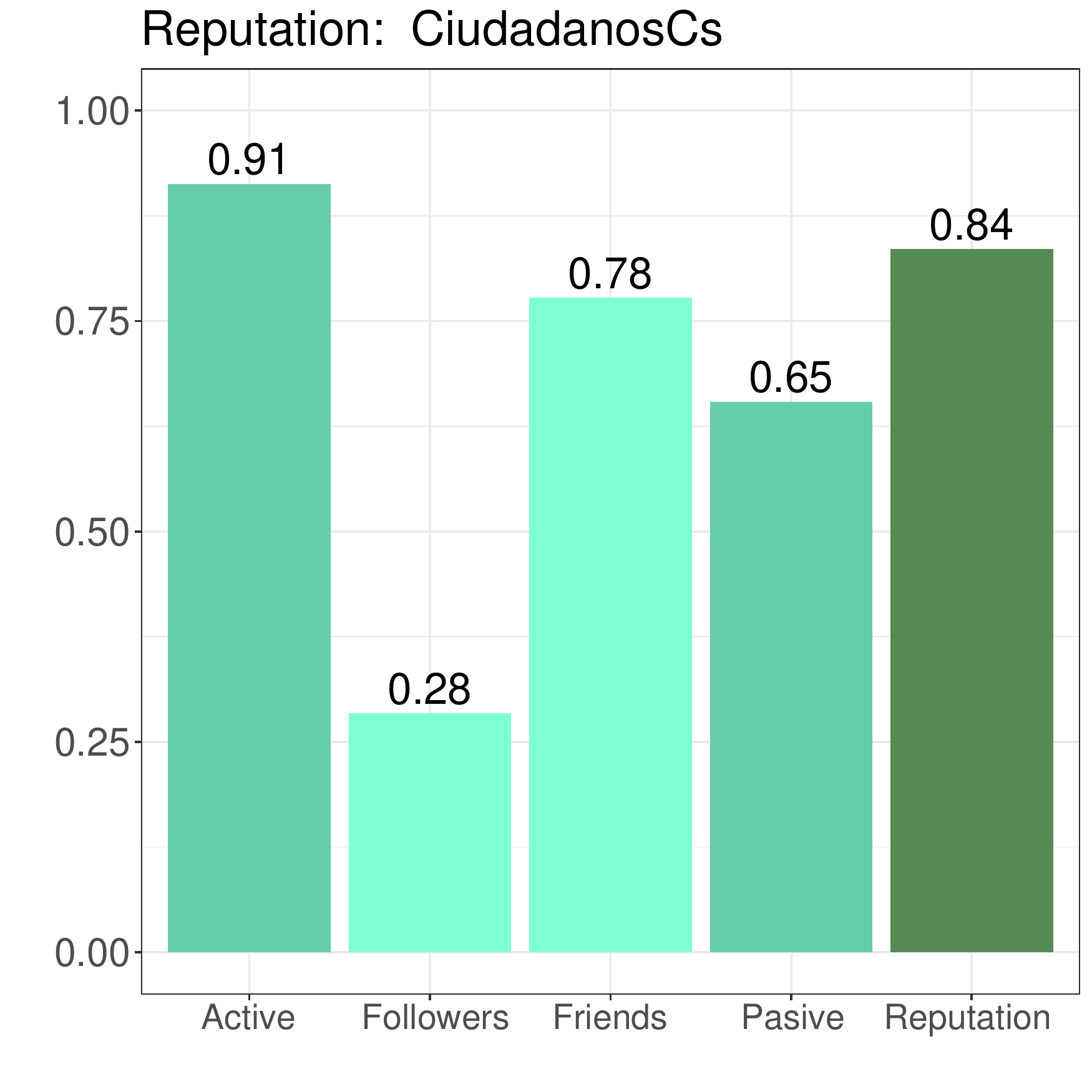}
    \end{subfigure}
    \begin{subfigure}{.19\textwidth}
      \centering
      \includegraphics[width=\linewidth]{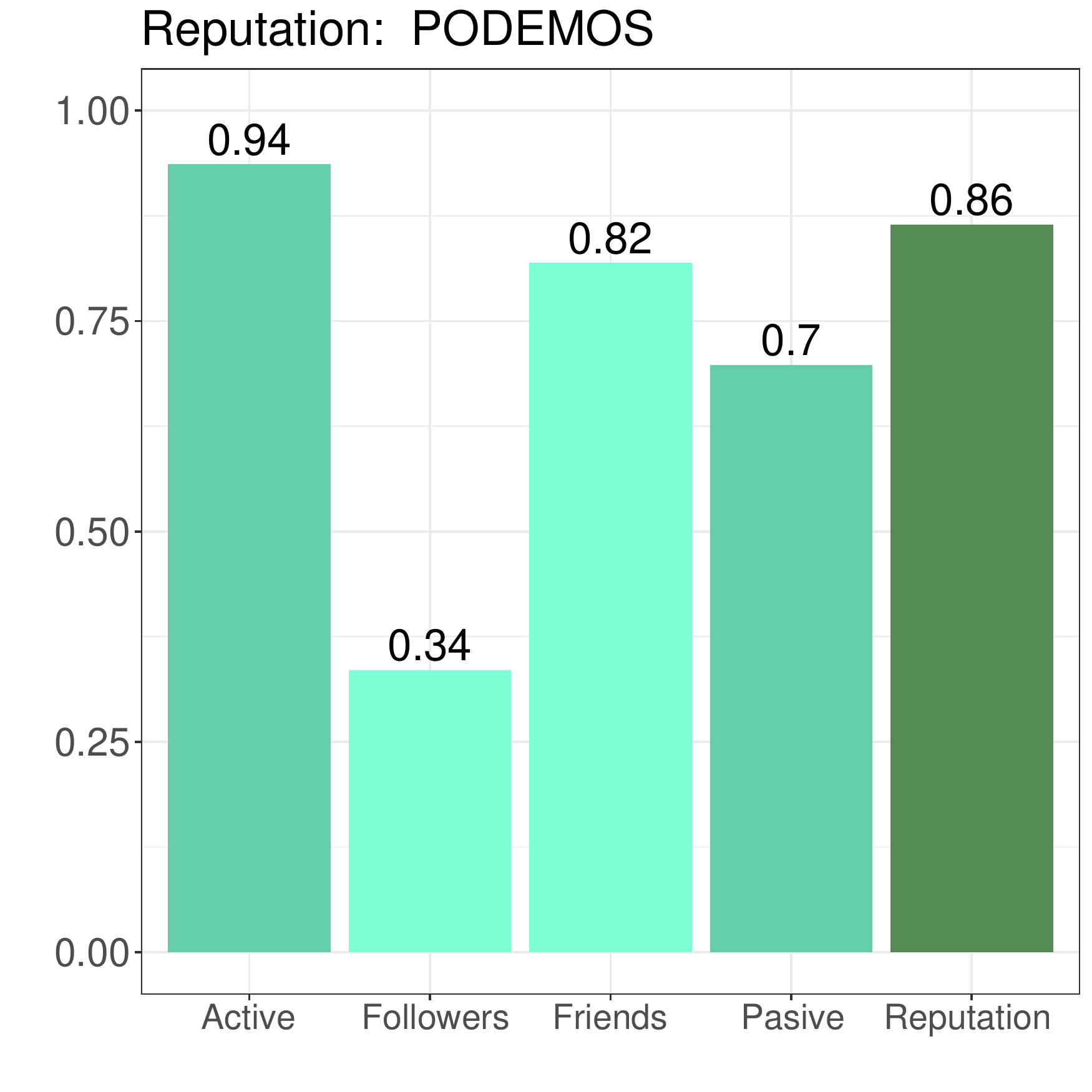} 
    \end{subfigure}
    \begin{subfigure}{.19\textwidth}
      \centering
      \includegraphics[width=\linewidth]{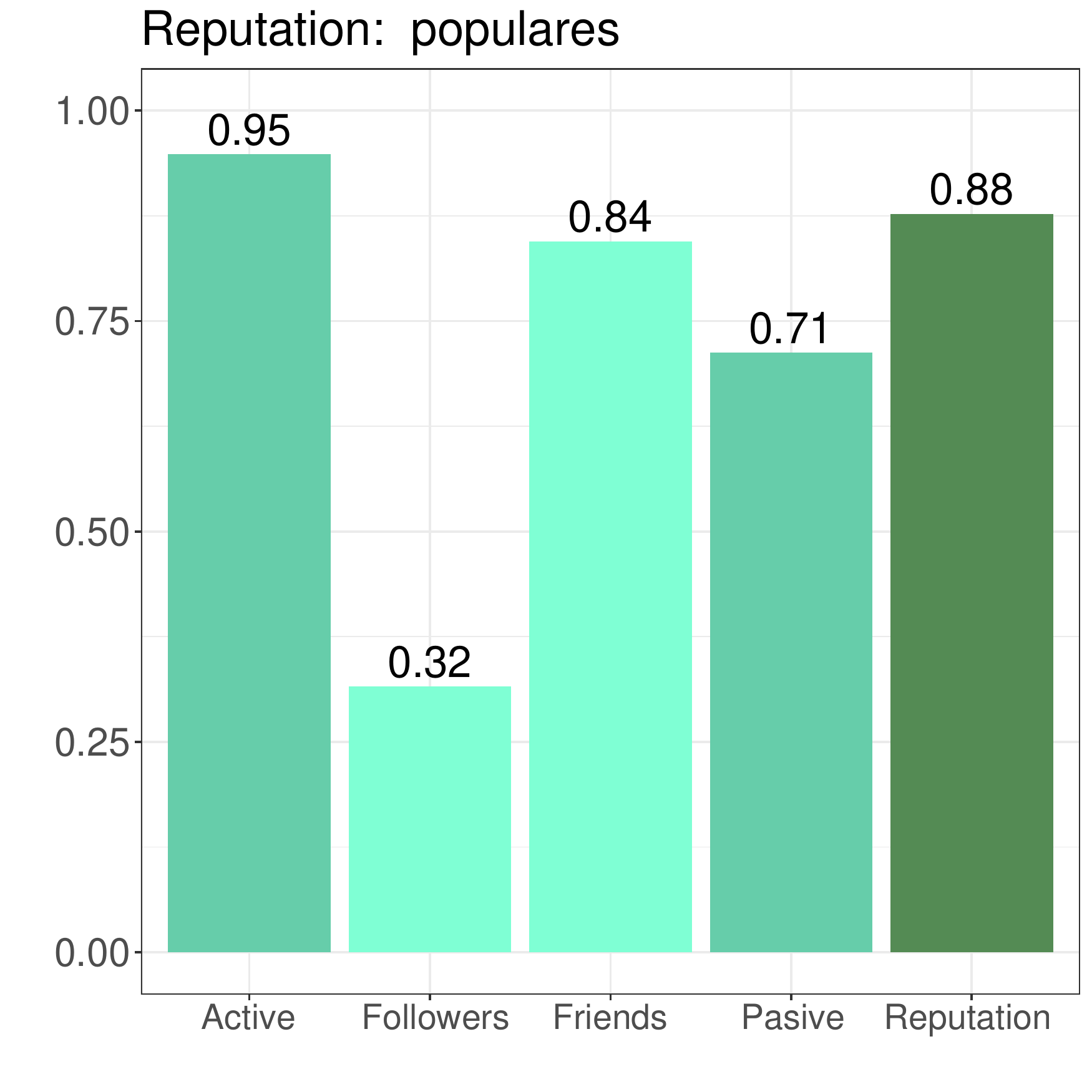}  
    \end{subfigure}
    \begin{subfigure}{.19\textwidth}
      \centering
      \includegraphics[width=\linewidth]{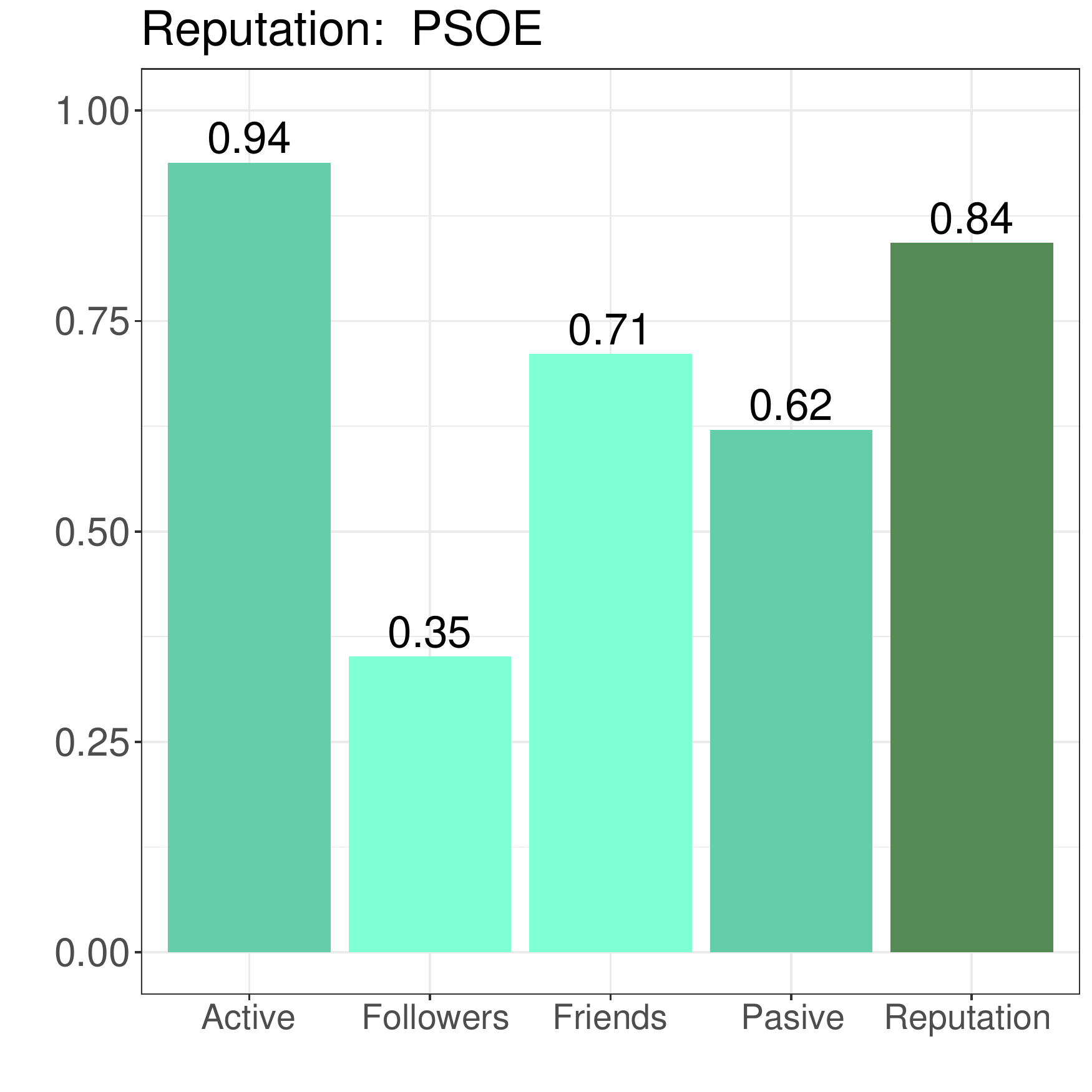}
    \end{subfigure}
    \begin{subfigure}{.19\textwidth}
      \centering
      \includegraphics[width=\linewidth]{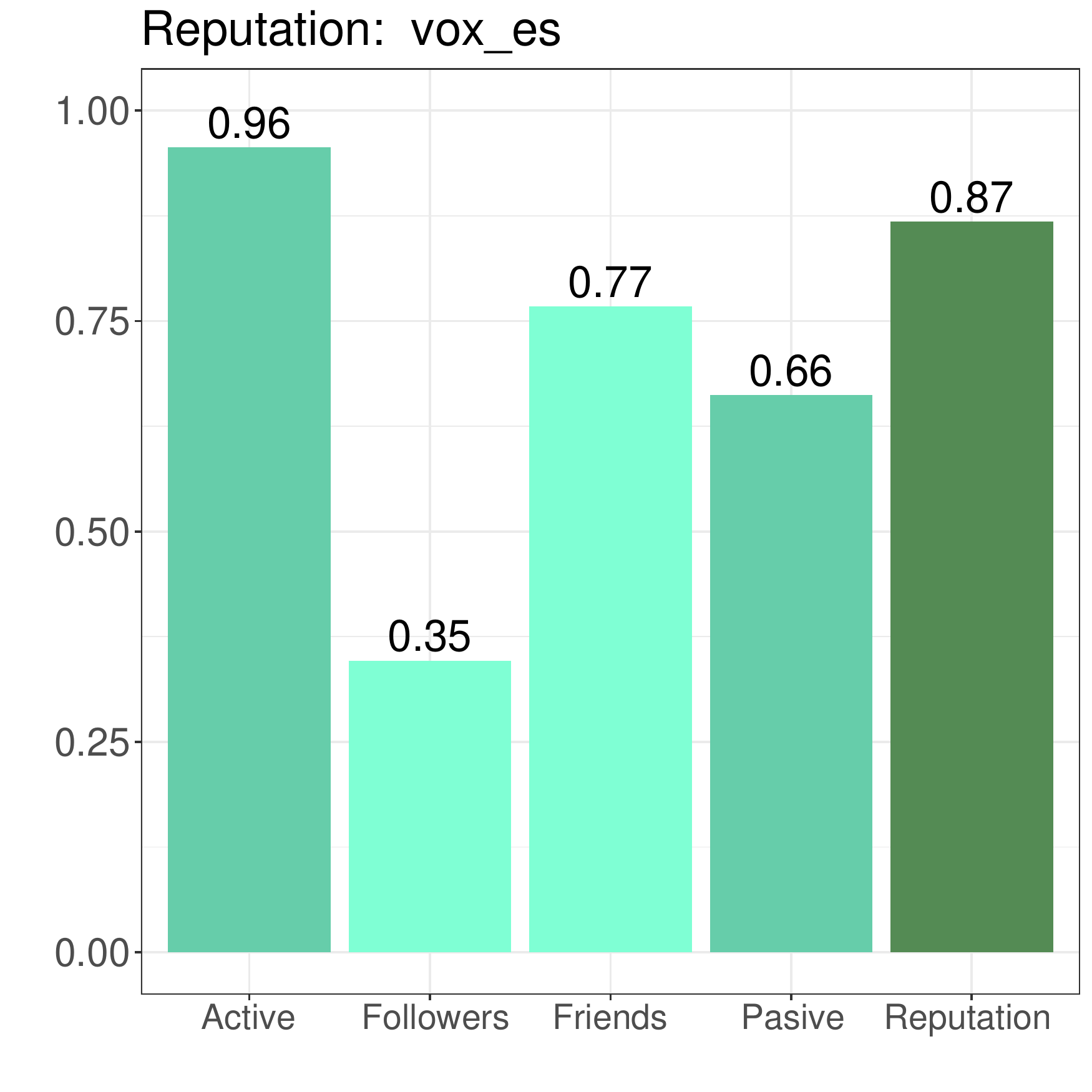}
    \end{subfigure}
    \caption{Reputation scores for our sample: bots (up) vs main parties (down). The color scale selected groups the indicators that are composed: ($R_P^I,R_P^O$), ($R_A,R_P$) and $R$.}
\label{fig:reputation}
\end{figure}

In order to check detailed information about each account, tables \ref{table:car_score_ejemplo} and \ref{table:reputacion} contains the values returned per each category. For comparisons among two groups (Bots and Parties), figure \ref{fig:reputation} contains the reputation vectors $\left(R_A,R_P, R_P^A, R_P^S, R\right)$ as a bar plot. Then, we can illustrate the same distribution of values per each group.

\begin{table}
\caption{Account features and active reputation of the sample. Bold values indicate the maximal of the column.} 
\centering
\begin{tabular}{p{0.20\textwidth}
                p{0.1\textwidth}p{0.1\textwidth}
                p{0.1\textwidth}p{0.1\textwidth}
                p{0.1\textwidth}p{0.05\textwidth}}
  \toprule
   Account     & User          & Cont          & Sent          & Temp          & Bot          & $R_A$ \\ 
  \midrule
  Bot\_5       & 0.02          & 0.04          & 0.13          & 0.04          & 0.02          & 0.04 \\ 
  Bot\_4       & 0.04          & 0.13          & 0.06          & 0.09          & 0.02          & 0.05 \\ 
  Bot\_2       & 0.04          & 0.30          & 0.05          & 0.09          & 0.02          & 0.06 \\ 
  Bot\_3       & 0.03          & 0.10          & 0.15          & 0.07          & 0.04          & 0.06 \\ 
  Bot\_1       & 0.04          & 0.29          & 0.87          & 0.09          & 0.02          & 0.11 \\
  CiudadanosCs & 0.93          & 0.96          & 0.95          & 0.87          & 0.86          & 0.91 \\ 
  PODEMOS      & 0.97          & 0.93          & 0.95          & 0.87          & 0.96          & 0.94 \\ 
  PSOE         & 0.94          & \textbf{0.97} & \textbf{0.96} & 0.91          & 0.92          & 0.94 \\
  populares    & 0.97          & 0.93          & 0.95          & \textbf{0.94} & 0.95          & 0.95 \\ 
  vox\_es      & \textbf{0.98} & 0.95          & 0.95          & 0.93          & \textbf{0.97} & \textbf{0.96} \\ 
  \bottomrule
\end{tabular}
\label{table:car_score_ejemplo}
\end{table}

\begin{table}
\centering
\caption{Reputation scores in its five versions of the sample. Bold values indicate the maximal of the column.} 
\begin{tabular}{p{0.20\textwidth}p{0.13\textwidth}
                p{0.13\textwidth}p{0.13\textwidth}
                p{0.13\textwidth}p{0.06\textwidth}}
  \toprule
  Account      & $R_A$           & $R_P^O$         & $R_P^I$         & $R_P$           & $R$ \\ 
  \midrule
  Bot\_5       & 0.0384          & 0.4885          & 0.7702          & 0.6998          & 0.2368 \\ 
  Bot\_4       & 0.0542          & \textbf{0.5001} & 0.7961          & 0.7221          & 0.2545 \\ 
  Bot\_3       & 0.0639          & 0.2399          & 0.8656          & 0.7092          & 0.2575 \\ 
  Bot\_2       & 0.0606          & 0.2960          & \textbf{0.9067} & \textbf{0.7540} & 0.2686 \\ 
  Bot\_1       & 0.1118          & 0.3423          & 0.8862          & 0.7502          & 0.3033 \\ 
  CiudadanosCs & 0.9130          & 0.2837          & 0.7775          & 0.6540          & 0.8353 \\ 
  PSOE         & 0.9381          & 0.3513          & 0.7108          & 0.6209          & 0.8429 \\ 
  PODEMOS      & 0.9362          & 0.3352          & 0.8189          & 0.6980          & 0.8648 \\ 
  vox\_es      & \textbf{0.9561} & 0.3465          & 0.7673          & 0.6621          & 0.8679 \\ 
  populares    & 0.9481          & 0.3158          & 0.8446          & 0.7124          & \textbf{0.8774} \\ 
   \bottomrule
\end{tabular}
\label{table:reputacion}
\end{table}

\subsection{Messages graph}\label{subsec:messages_graph}

\begin{figure}[b!]
\centering
\begin{subfigure}{\textwidth}
  \centering
  % include first image
  \includegraphics[width=0.825
  \linewidth]{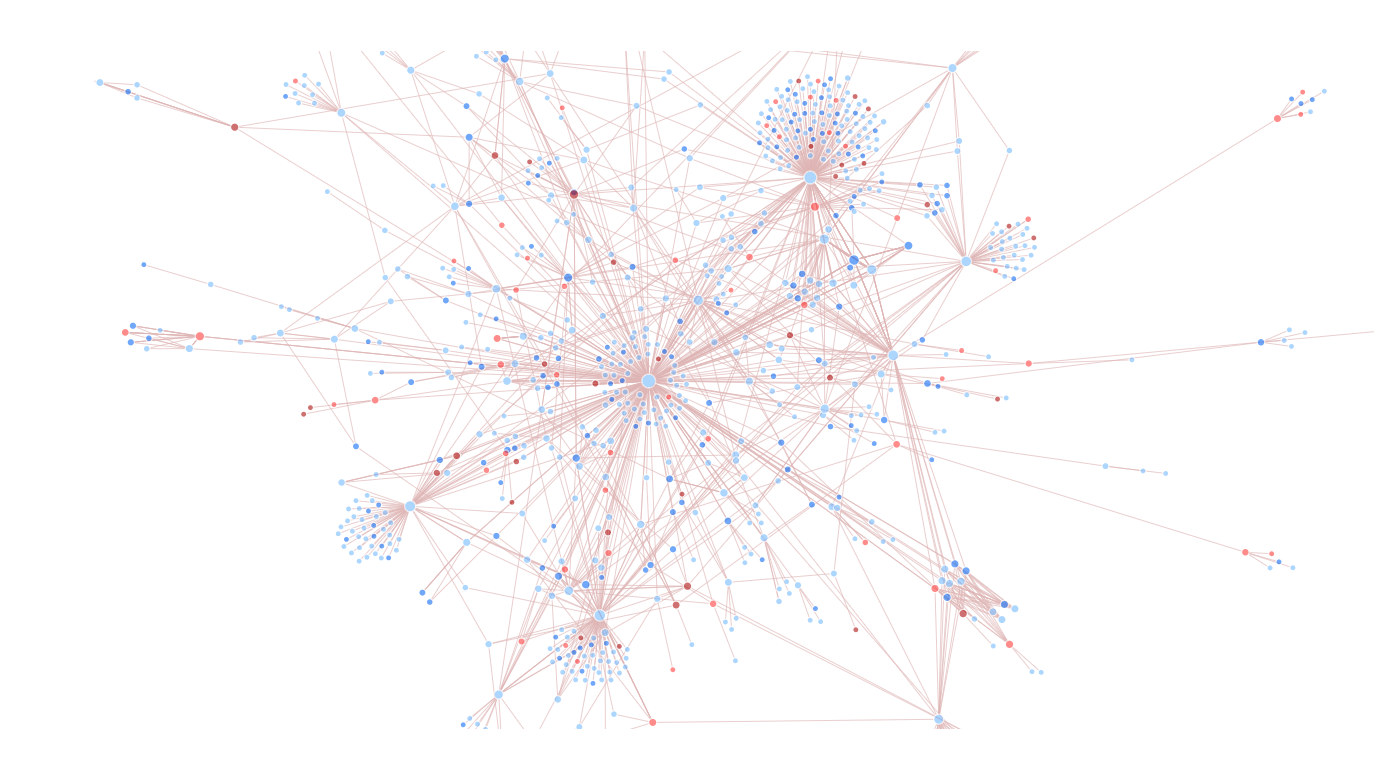}
\end{subfigure}

\begin{subfigure}{.33\textwidth}
  \centering
  \includegraphics[width=\linewidth]{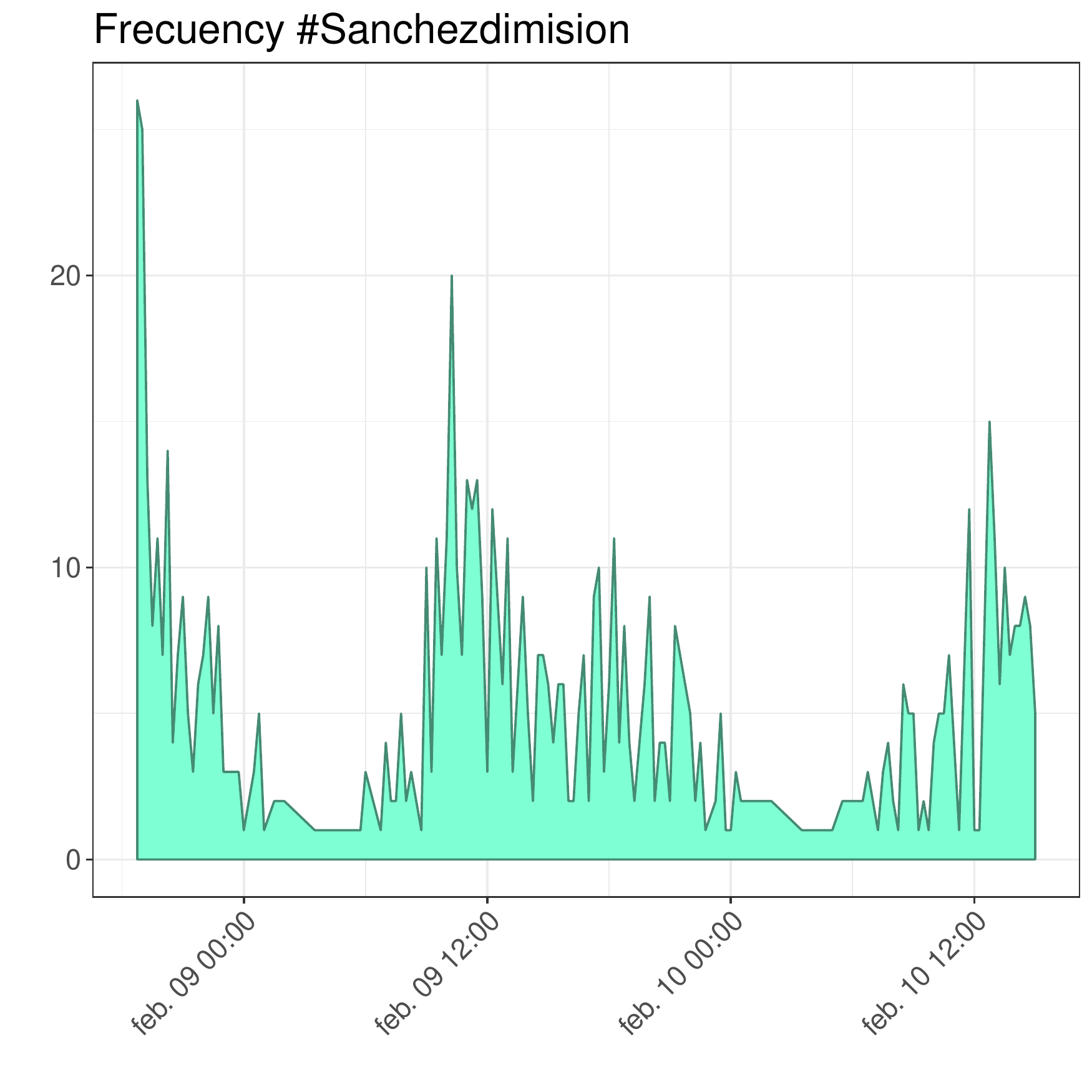}
\end{subfigure}
\begin{subfigure}{.33\textwidth}
  \centering
  \includegraphics[width=\linewidth]{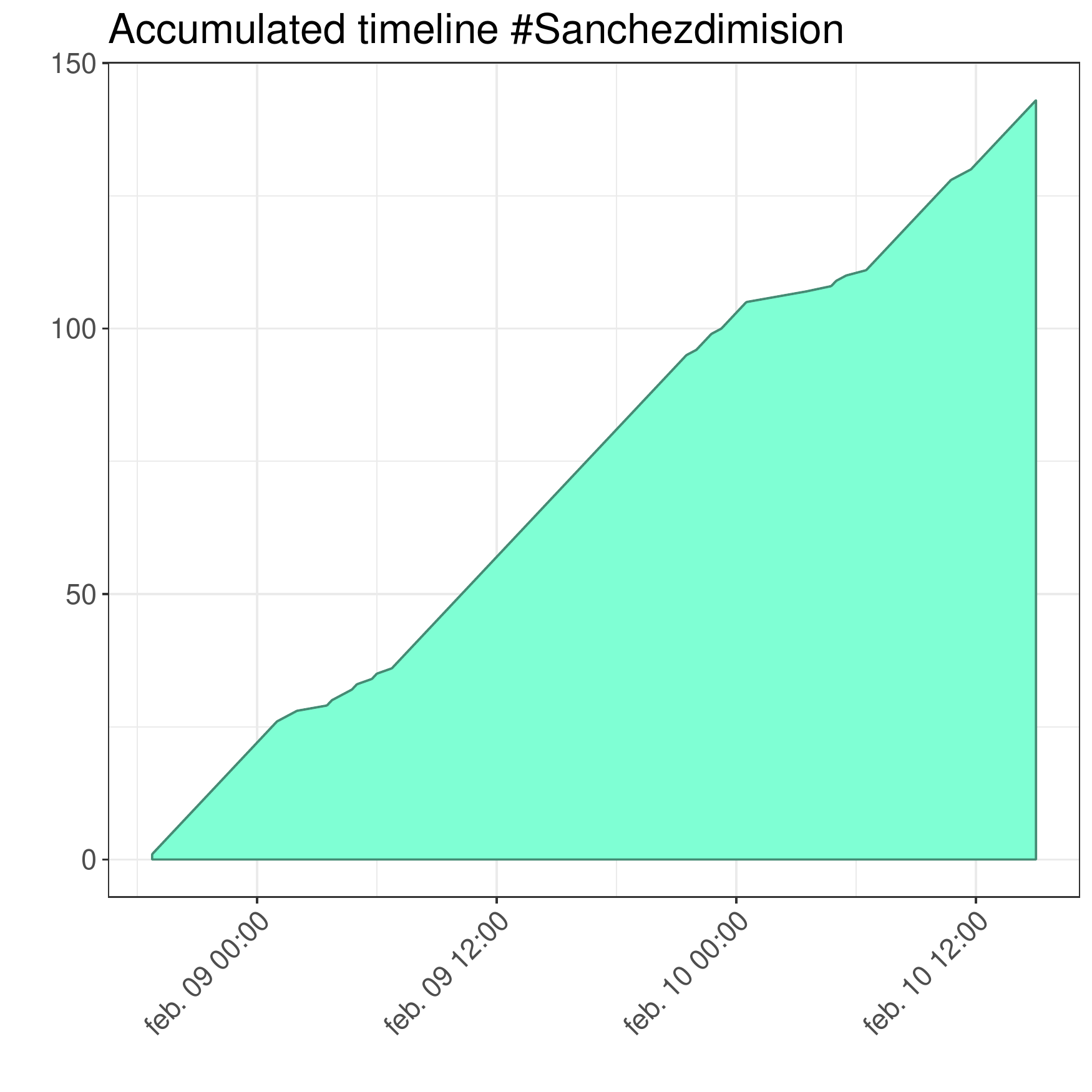} 
\end{subfigure}
\begin{subfigure}{.33\textwidth}
  \centering
  \includegraphics[width=\linewidth]{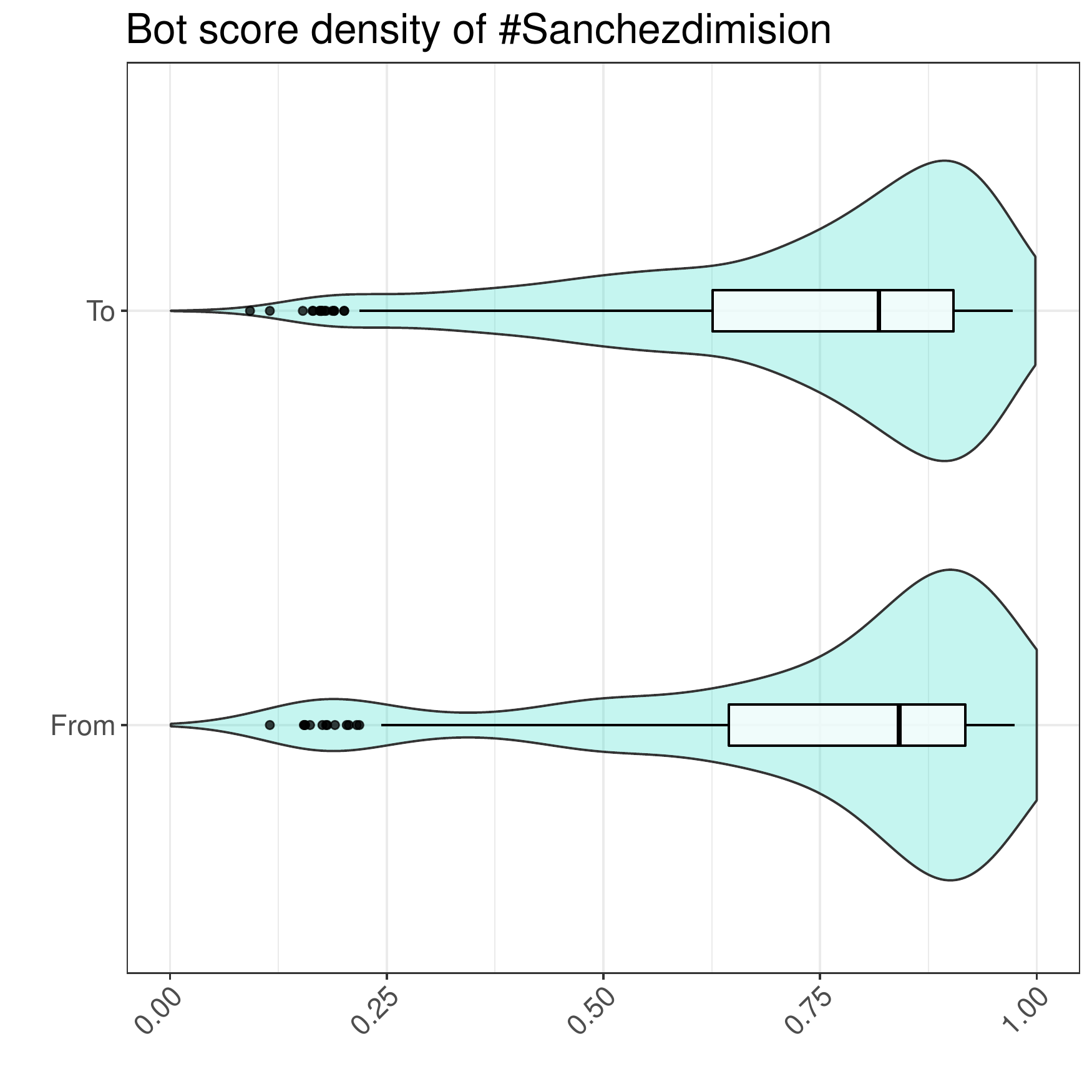}  
\end{subfigure}
\begin{subfigure}{.33\textwidth}
  \centering
  \includegraphics[width=\linewidth]{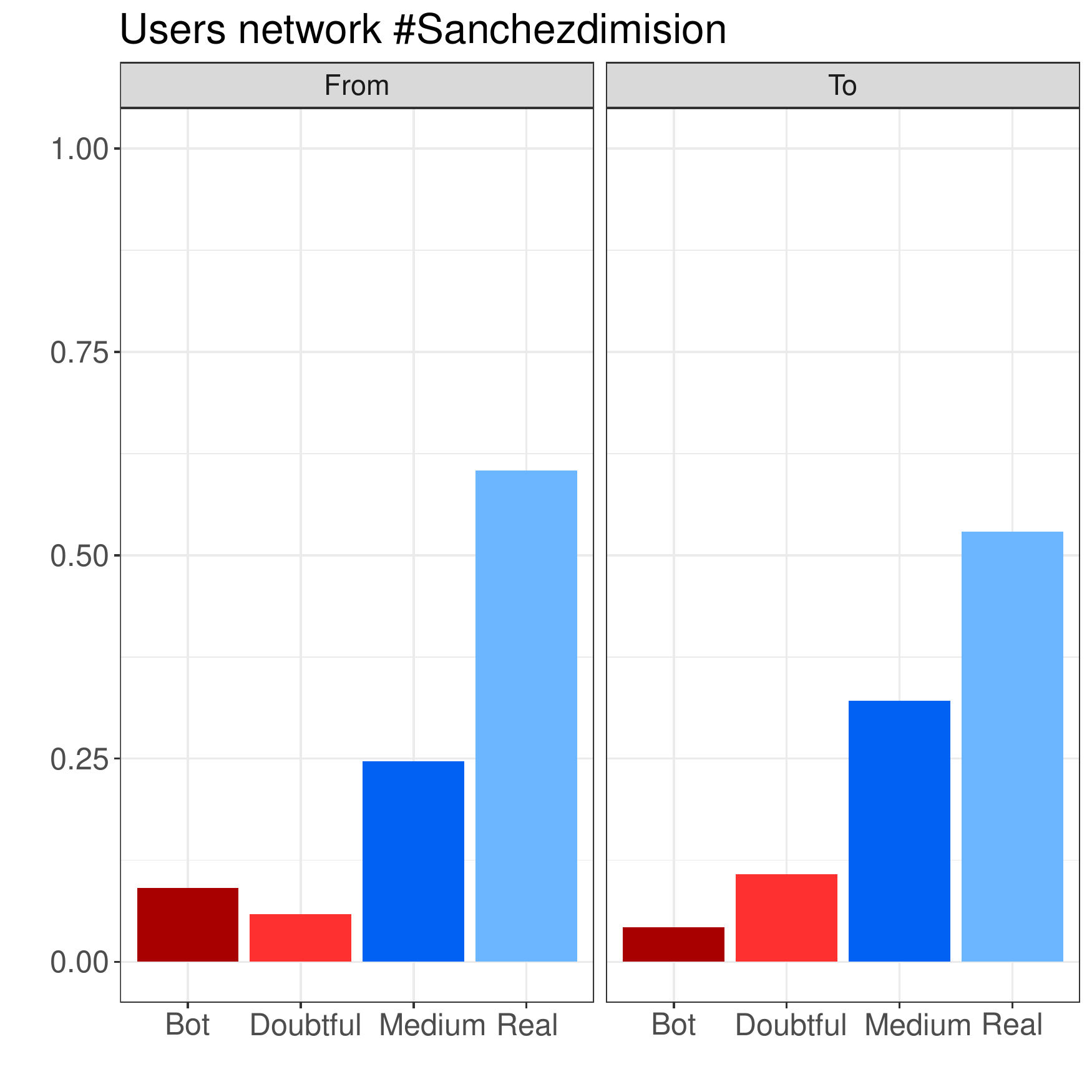}  
\end{subfigure}
\begin{subfigure}{.33\textwidth}
  \centering
  \includegraphics[width=\linewidth]{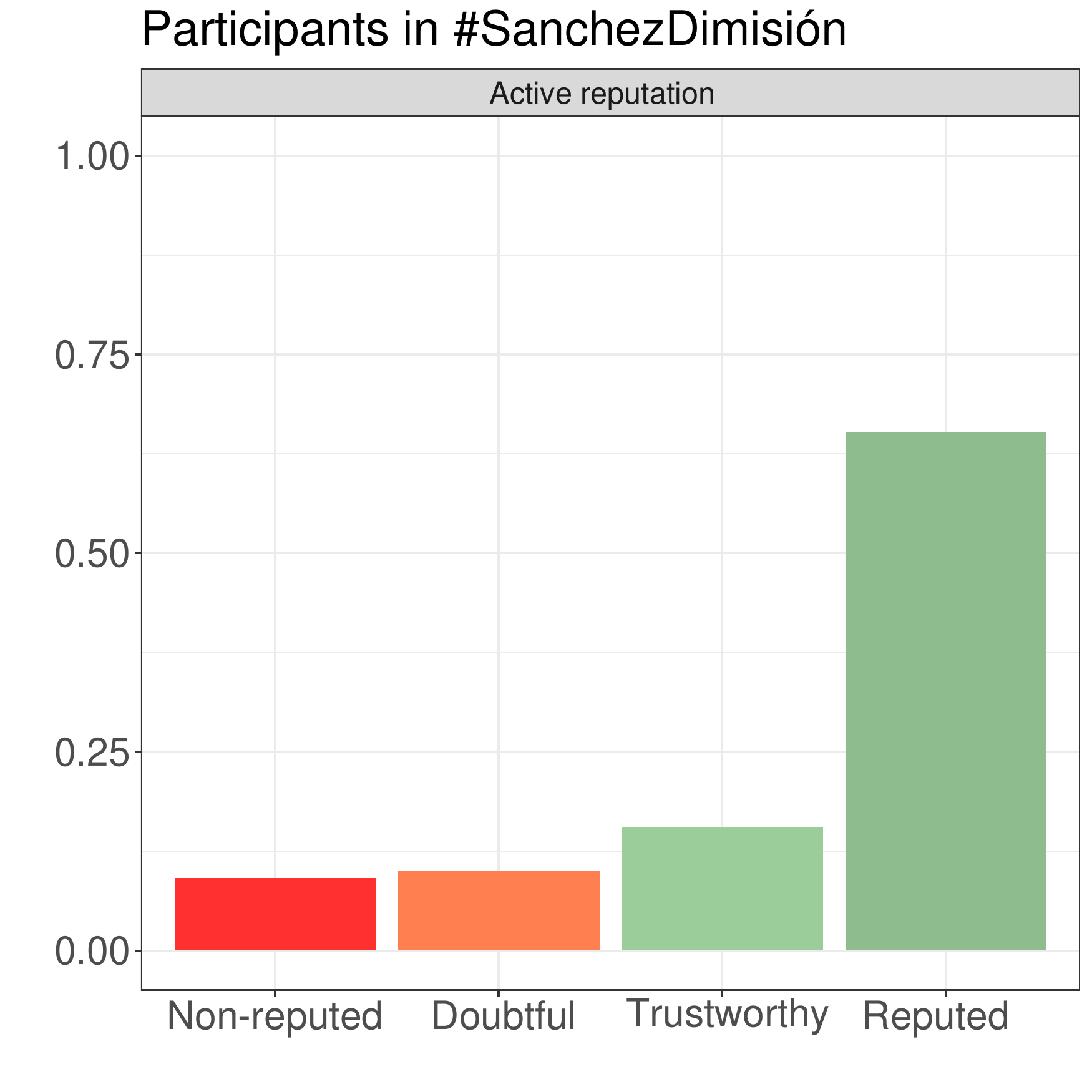}  
\end{subfigure}
\begin{subfigure}{.33\textwidth}
  \centering
  \includegraphics[width=\linewidth]{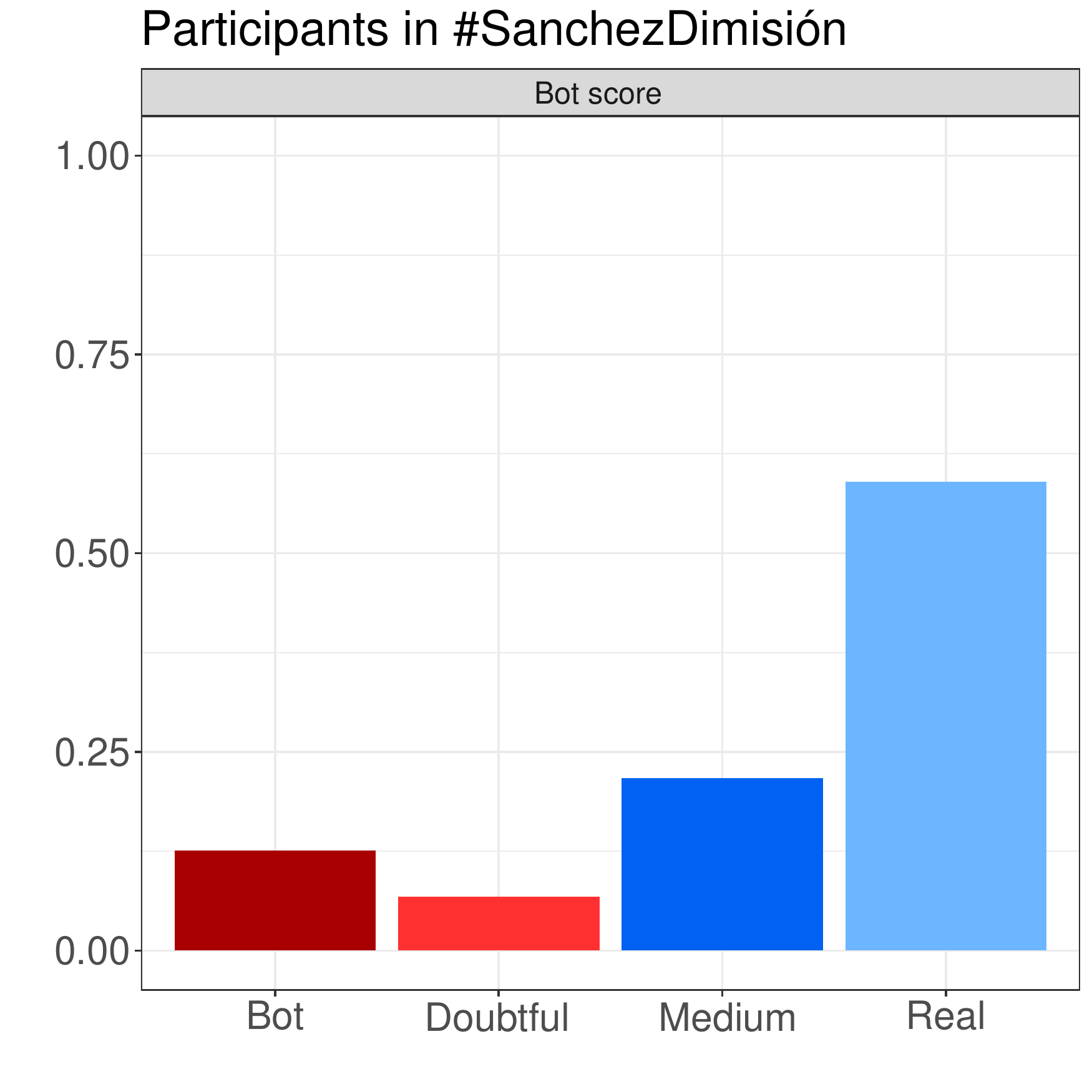}  
\end{subfigure}
\caption{Bot apparition and active reputation analysis of the timeline $\#$Sanchezdimision}
\label{fig:Sanchezdimision}
\end{figure}

The tree structure in a Twitter's timeline define its directions and its temporal apparition. One of the most controversial trends during the Spanish campaign was the $\#$SanchezDimisión hashtag. Figure \ref{fig:Sanchezdimision} shows a detailed explanation of the graph created at the time in which it is captured the temporal evolution together with the bot score and active reputation. Here the graph is $G_T$ based, where vertices represent the tweets posted by accounts. The definitions of the subsection \ref{subsec:graph_analysis} are used for in depth study of centrality and intermediation among users or tweets.

\section{Conclusions}\label{sec:conclusions}

The main goal of this paper is to propose a synthetic indicator that measures the reputation of Twitter accounts in order to analyze the information flow and spot malicious spreaders. With the results extracted from the social media, we can make decisions in uncertain scenarios.

In respect of the apparition, tracking, and behaviour of bots; we have shown that there exists a large number of fake accounts in the studied net. The tactic of bots is clear, they follow each other in order to gain centrality within graphs (either user or tweet based). It is worth mention that the definition of bot is stated in table \ref{table:baremo_bot}, hence such classification is biased in regard of our criteria.

Regarding the proposed indicators, our approach suitably reflects the concept of fake-real account. For the selected sample, the maximal active reputation ($R_A$) for bots is $0.1118$ and for verified political parties the minimal is $0.9130$, exposing significant evidences of the precision of our concept of reputation. Another interesting point is that bots have higher values for the passive reputation ($R_P$), especially for the friends net. We already expected that because their strategy is to be as close as possible to relevant accounts and events. Anyway, it is shown in figure \ref{fig:example} for the bot category box. For total reputation ($R$), we can see that the maximal total reputation for bots is $0.3033$ and for verified political parties the minimal is $0.8353$, giving a lower range than active reputation. That is crucial and it is well designed with our approach since we have followed the idea that every account must control both their activity and the net in which it is involved.

So far we have not encountered temporal disorders in our samples. For instance, the activity showed in figure \ref{fig:Sanchezdimision}, is regularly distributed over the time and there is not too much publications between 00:00 and 6:00. Sometimes it is studied when there exists evidences of external influence over a topic.

Finally, we can conclude by saying that the Spanish presidential election on November of 2019 was partially adulterated by the presence of fake accounts and non reputed users in Twitter. For the case before mentioned in subsection \ref{subsec:messages_graph}, the last row of the figure \ref{fig:Sanchezdimision} indicates that the timeline of $\#$SanchezDimisión contains at least $12.56\%$ of bot accounts and $8.65\%$ of non-reputed users. Another evidence, slightly documented, is that bots are more likely to share content ($5.13\%$ more) than create it. Therefore, their main role is to spread content.

\bibliographystyle{unsrtnat}
\bibliography{export}  %%% Uncomment this line and comment out the ``thebibliography'' section below to use the external .bib file (using bibtex) .





% \end{thebibliography}

\end{document}